\documentstyle[11pt,newpasp,twoside]{article}
\def\edcomment#1{\iffalse\marginpar{\raggedright\sl#1\/}\else\relax\fi}
\reversemarginpar

\begin{document}
\title{Problems with the Current Cosmological Paradigm}
 \author{T. Shanks}
\affil{Department of Physics, University of Durham,
South Road, Durham DH1 3LE, England}

\begin{abstract} We note that despite the apparent support for the
$\Lambda$CDM model from the acoustic peaks of the CMB power spectrum
and the SNIa Hubble diagram, the standard cosmological model continues
to face several fundamental problems. First, the model continues to
depend wholly on two pieces of undiscovered physics, namely dark
energy and cold dark matter. Then, the implied dark energy density is
so small that it is unstable to quantum correction and its size is
fine-tuned to the almost impossible level of one part in
$\approx10^{102}$; it is also difficult to explain the coincidence
between the dark energy, dark matter and baryon densities at the
present day. Moreover, any model with a positive $\Lambda$ also
creates fundamental difficulties for superstring theories of quantum
gravity.  We also review the significant number of astrophysical
observations which are now in contradiction with the $\Lambda$CDM
model. On the grounds that the SNIa Hubble diagram is prone to
evolutionary corrections and also that the CMB power spectrum may be
contaminated by the effects of foreground ionised gas, we argue that
the existence of such systematics could still allow more satisfactory,
alternative, models to appear. We suggest that if
$H_0\la50\,$kms$^{-1}$Mpc$^{-1}$ then a simpler, inflationary model
with $\Omega_{baryon}=1$ might be allowed with no need for dark energy
or cold dark matter. We note that the clear scale error between HST
Cepheid and Tully-Fisher galaxy distances and also potential
metallicity dependencies for both the Cepheid P-L relation and the
SNIa Hubble diagram may mean that such a low value of $H_0$ cannot yet
be ruled out.

\end{abstract}

\section{Introduction}

It is a recurrent recent theme that we live in a `New Age of Precision
Cosmology' to the point where we may even be witnessing `the end of
cosmology'. These views  are prompted by the cosmic microwave background
anisotropy results from Boomerang and WMAP (Netterfield et al., 2002,
Hinshaw et al., 2003) on the one hand and the SNIa Hubble Diagram
results on the other (Riess et al., 1998, Perlmutter et al., 1999).
These results both appear to indicate that the Universe is dominated by
Cold Dark Matter and Dark Energy. But both fundamental and astrophysical
problems for $\Lambda$CDM remain. These are significant enough to
suggest that continued inspection of the current cosmological data for
ways out of the current `concordance' model may still be worthwhile.
Here, after considering the fundamental problem areas for the standard
model, we shall look at the CMB and SNIa results which are the main 
observational pillars of the model and suggest that they may be more
susceptible to  systematic error than currently emphasised. This shall
prompt us to look at alternative models which drop the assumption of
either cold dark matter or dark energy or both.

\section{A new age of precision cosmology?}

The idea that the age of precision cosmology has dawned, is based on the
Boomerang and WMAP CMB anisotropy experiments' detections of the
first acoustic Doppler peak at $l=220$  ($\approx$1-2 deg). Such a
large spatial scale for the first peak is expected  in a spatially flat,
CDM Universe. The confirmation of the Boomerang results by the WMAP
experiment has removed any doubt as to the observational reality of this
detection. This observation is complemented by the evidence for an
accelerated expansion seen in the SNIa Hubble Diagram. Jointly, these
two observations appear to require a zero spatial curvature Universe
with $\Omega_\Lambda=0.7$ and $\Omega_m=0.3$.

Although the argument for the standard model has undoubtedly been
strengthened by the above two observations, fundamental problems still
remain. For example, the standard $\Lambda$CDM model still relies on two
pieces of undiscovered physics! The first is the CDM particle for which
there is still no laboratory detection, some twenty years after it was
first proposed (Blumenthal et al., 1982, Bond, Szalay \& Turner, 1982, Peebles, 1982). For the
optimists, the search for the CDM particle is likened to the search for
the neutrino in the 1930's but for the pessimists the situation may be
more like the search for the electro-magnetic ether at the end of the
19th Century. The second piece of undiscovered physics is dark energy.
The invoking of dark energy also makes $\Lambda$CDM complicated and
fine-tuned. There are two separate fine-tuning problems associated with
dark energy, at least when it is represented as a
cosmological constant. First, the vacuum energy term is small;
after inflation it is only one part in $10^{102}$ of the energy density
in  radiation. This small size means that the dark energy is unstable to
quantum correction (e.g. Dvali, Gruzinov \& Zaldarriaga, 2003). Second,
there is the coincidence that it is only relatively close to the present
day where $\Omega_\Lambda\approx\Omega_{m}$; there seems no clear reason
why the present day should have this special status. Even for those who
dislike fine-tuning arguments, to start with one fine tuning (flatness)
problem and end up with several seems circular!

Several solutions have been proposed to solve the $\Lambda$ fine-tuning
problems. For example, quintessence is the name given to the dark vacuum
energy when it takes the form of a scalar field slowly rolling down a
potential, usually from an initially high value,  until the present day
(Wetterich, 1988; Peebles \& Ratra, 1988). Indeed, the initial value can be
comparable to the radiation energy density after inflation, thus
addressing the first $\Lambda$ fine-tuning problem. However, it offers
no solution to the second $\Lambda$ fine-tuning problem of the
coincidence with the matter energy-density at the present day. 

Another solution is represented by the aptly-named Cardassian model
(Chung \& Freese, 2000, Freese \& Lewis, 2002) where an extra term is
added to the Friedmann equation so that $H^2=A\rho+B\rho^n$, with
$n<2/3$.  (A related model is the brane-induced gravity model of Dvali,
Gabadadze \& Porrati, 2000). The extra power-law term could arise from
gravitational effects caused by  embedding the Universe as a 3(+1)-D
brane in a higher dimensional entity. Here the accelerated expansion
arises from the extra term associated with the matter density, $\rho$.
This has the benefit of removing the need for dark vacuum energy and
even cold dark matter and so could be said to reduce the dependence of
the model on undiscovered physics. The removal of dark energy again
addresses the first $\Lambda$ problem but the second problem of why the
acceleration only starts to dominate at the present day is again left
unaddressed.

We note that a further problem has appeared for any model with a
positive cosmological constant in that superstring theories of quantum
gravity with compactified extra spatial dimensions are much more viable
in models where $\Lambda<0$ (Anti-de Sitter space) than in cosmologies
where $\Lambda>0$ (Banks, 2000, Witten, 2001, Deffayet, Dvali \&
Gabadadze, 2002). Although solutions have been suggested to this problem
they appear highly contrived (Kachru et al., 2003). Thus there are
many fundamental problems involved with the size and sign of the dark
energy density required by the standard model. So unnatural does a
small, positive cosmological constant appear to be that several authors
have resorted to invoking the anthropic principle as the most likely
hope for an explanation (Efstathiou, 1995, Martel, Shapiro \& Weinberg,
1998).

Even without dark energy, further fundamental problems are inherent in
any model based on CDM. First, as noted by Peebles (1984), any CDM model
has some fine-tuning since $\Omega_{CDM}\approx\Omega_{baryon}$.
Attempts have previously been made to explain this coincidence if the
cold dark matter particle has approximately the mass of the proton (Turner
\& Carr, 1986, priv. comm.), but the accelerator lower limit on the mass of the
neutralino, for example, is now an order of magnitude higher than
this. Second, baryonic dark matter is needed anyway since
nucleosynthesis  implies that
$\Omega_{baryon}\approx10\times\Omega_{star}$. The baryonic  candidate
for the $\Omega_0\approx0.1$ dark matter may then be a contender also
for the $\Omega=1$ dark matter candidate (see Section 5 below). Third,
the dark matter in the Coma cluster has a significant baryon component
with $\approx$20\% of the virial mass of Coma  now well known to be hot
X-ray gas (Lea et al., 1973). The discovery of substantial amounts of
X-ray gas in clusters such as Coma has reduced the Coma mass-to-light
ratio from M/L$\approx$60-600 to M/L$\approx$5. If the Coma `missing
mass' problem is only at the level of M/L$\approx$5 then it may be
considered less plausible to invoke a cosmological density of exotic
particles than if M/L$\approx$60-600! If Zwicky had known about the
X-ray gas in Coma, the question is whether he would have been inclined
to introduce the term `missing mass' at all!

\section{Astrophysical problems for $\Lambda$CDM}

There are several other problems for the $\Lambda$CDM model which might
be classed more observational or astrophysical than fundamental. First,
the mass profiles of low surface brightness galaxies appear to be less
sharply peaked than predicted by CDM models (Moore et al, 1999a).
Second, the large numbers of sub-haloes predicted in galaxy haloes may
make spiral disks subject to tidal disruption on timescales of less than
a Gigayear (Moore et al., 1999b). Third, the observed galaxy luminosity
function is much flatter than the mass distribution predicted by CDM;
attempts to suppress star-formation by invoking significant feedback in
low-mass haloes appear to create further problems at higher masses
(Benson et al., 2003). Fourth, the slope of the galaxy correlation
function is  flatter than predicted by $\Lambda$CDM, suggesting that the
galaxy distribution must be anti-biased on scales $r<1h^{-1}$Mpc. This
means that a simple  high peaks bias model is disallowed (Colin et al.,
1999) - although this is not a problem in principle, it does mean that
the bias model has to be relatively complicated. Fifth, the $L_X-T$
relation for galaxy clusters is not scale-free as predicted by
hierarchical models (Lloyd-Davies et al. 2000). Some attempts have been
made to fix things by suggesting that at small-scales, entropy might be
increased by shocks created during the process of galaxy formation (Voit
et al., 2003). However, the simpler explanation is that it is the mass
distribution that is not scale free and this would represent a
fundamental argument against hierarchical models such as $\Lambda$CDM.

Of course, any evidence that $\Omega_m\approx1$ could be taken as
evidence against the standard $\Lambda$CDM model which requires
$\Omega_m\approx0.3$. One such piece of evidence comes from the lensing
of background QSOs in the 2dF QSO redshift survey by foreground galaxy
groups and clusters (Croom \& Shanks 1999, Myers et al., 2003). These
authors find a high lensing mass per cluster which leads to a 2$\sigma$
rejection of the $\Omega_m=0.3$ model.

\begin{figure}
\plotfiddle{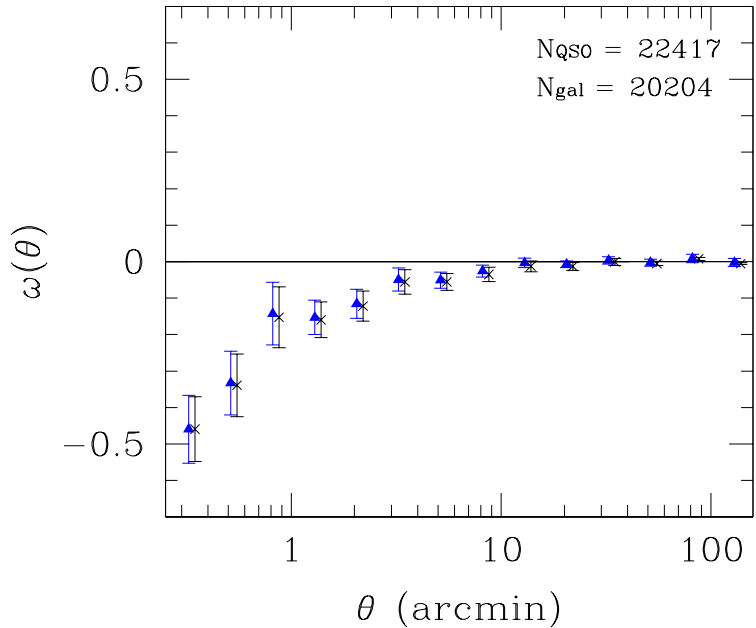}{3in}{0}{55}{55}{-180}{-90}
\caption{The 2-D spatial cross-correlation between QSOs and foreground
APM/SDSS selected galaxy groups and clusters from Myers et al. (2003).
The anti-correlation is the result expected if the foreground clusters
are lensing the background QSOs in a high-density, $\Omega_m\approx1$
Universe. The similarity of the results shown by the squares and
triangles show the anti-correlation is robust to whether the search is
made for QSOs around clusters or vice-versa. } 
\end{figure}

Evidence for $\Omega_m\approx1$ even arises from the space abundances of
galaxy clusters (Eke et al., 1998, Vauclair et al., 2003). The evolution
of clusters is often quoted as vital evidence for the concordance model.
But many of these estimates seem remarkably close to $\Omega_m=1$.
Vauclair et al. claim that the data support $0.8<\Omega_m<1$. The best
estimate of Eke et al. is $\Omega_m =0.45\pm0.25$. Even in the latter
case, it might be recalled Guth (1981) argued that the $\Omega_m>0.01$
lower limit from nucleosynthesis left $\Omega_m$ embarassingly close to
unity and  now even the estimate of Eke et al. lies within a factor of
two of the Einstein-de Sitter value.

\section{Escape routes: SNIa evolution $+$ CMB foreground contamination}

Given this collection of fundamental and astrophysical problems, it is
worthwhile considering if there are any  escape routes from the
observations that underpin standard model. The escape route from the
SNIa Hubble diagram is certainly clear; there is the obvious possibility
that the SNIa maximum luminosity evolves with look-back time in a way
that is not detectable in the SNIa spectra. The SNIa are $\approx$0.5mag
fainter at z$\approx$0.5 if $\Omega_\Lambda=0.7$ and $\Omega_m=0.3$ than
in the Einstein-de Sitter case. Quite natural evolutionary mechanisms
for SNIa certainly exist. For example, the metallicity of the SNIa
progenitor stars at high redshift are likely to be lower than they are
locally. Also the C/O abundance ratio of the White Dwarf will change as
it awaits the accretion of mass which will trigger the explosion. These
evolutionary corrections are likely to be comparable to the above effect
of $q_0$ (Hoeflich et al., 2000).

In the case of the CMB power spectrum, the main escape route here is
likely to be the CMB foregrounds. Although there are now quite good
constraints from the CMB spectral index on contamination from Galactic
synchrotron and dust, the WMAP results have suggested two other sources
of foreground contamination. The excess TE polarisation detected by WMAP
at large angular scales is interpreted as strong evidence for an early
epoch of reionisation at $10<z<20$ with optical depth, $\tau\approx0.17$
(Kogut et al., 2003). Homogeneous reionisation with this optical depth
reduces the amplitude of the temperature power spectrum peaks by
$\approx$30\%. Inhomogeneous reionisation could also alter the peak
shapes. Although this is expected only to affect the smaller peaks, the
large-scale peaks could also be affected, depending on the model and the
details of the reionisation process.

\begin{figure}
\plotfiddle{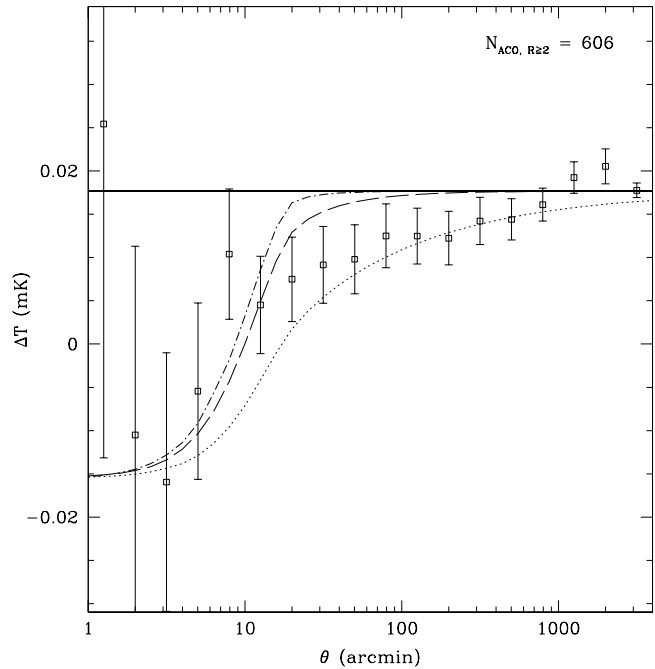}{3in}{0}{45}{45}{-140}{-75}
\caption{ Cross-correlation of WMAP 94$\,$GHz W band data with ACO $R\geq 2$
Abell clusters for combined ACO $|b|>40\,$deg N$+$S samples. The dashed,
dotted and dot-dash lines are isothermal models for the SZ decrement 
as presented by Myers et al. (2004).} 

\end{figure}

Another source of foreground contamination could be due to the SZ effect.
Myers et al. (2004) have cross-correlated the Abell $R\ge2$,
$|b|>40\,$deg, clusters with the WMAP 94GHz W band data and found
signifiant anti-correlation which they interpret as due to the SZ
effect. Similar signals were found in the groups and clusters detected
in the APM and 2MASS catalogues.  Interestingly, they found that in
the case of the rich clusters the anti-correlation appeared to extend
to scales larger than the $12.'6$ W-band beam size, out to scales of
$\approx 1\,$deg ($\approx 5$h$^{-1}$Mpc) which could be caused by
ionised supercluster gas. Although the significance of the
extended signal is lower than on the beam-size, if it is real then there
could be important implications. In particular, there could be a
significant SZ contribution to even the first peak of the power
spectrum on $\approx1-2\,$deg scales. Thus on grounds of both the
ionised gas at the epoch of reionisation at $z\approx15$ and the hot
gas in clusters at lower redshift, the CMB signal may have come
through more foreground `traffic' than previously expected and the
resulting contamination may have seriously compromised its primordial
signal.

\section{$H_0$ route to a  simpler model}

Given that the quintessence and Cardassian modifications to the
standard model only represent partial solutions to the problems of
dark energy and dark matter we next consider a previously suggested
route via $H_0$ to a simpler model. Shanks (1985, 1991, 1999, 2000,
2002) suggested that if $H_0\la30$kms$^{-1}$Mpc$^{-1}$ then there
might be no need to introduce either dark matter or dark energy. With
a low value of $H_0$, an inflationary model with $\Omega_{baryon}$=1
is then better placed to escape the baryon nucleosynthesis constraint,
since $\Omega_0=\rho_0/\rho_c$ and $\rho_c=3{H_0}^2/8\pi
G$. Simultaneously, the low value of H$_0$ means that the X-ray gas in
the Coma cluster increases towards the Coma virial mass, since
$M_{gas}/M_{virial}\propto{H_0}^{-1.5}$. Finally, the lifetime of an
Einstein-de Sitter Universe increases as $1/H_0$ to become compatible
with the ages of the oldest stars. Given the historical uncertainty
there has been in observational estimates of H$_0$, the potential
simplification in cosmology that this very simple model offers,
removing the need for dark energy and cold dark matter, provides clear
motivation to continue to investigate the distance scale and
Hubble's constant.

The value of Hubble's constant has been notoriously difficult to
estimate. prior to the opening of the Palomar 5-m telescope in 1950, 
Hubble's value was $H_0\approx500\,$kms$^{-1}$Mpc$^{-1}$. Since then,
estimates of $H_0$ have moved down to
$H_0\approx70\,$kms$^{-1}$Mpc$^{-1}$. We now argue that the value of
$H_0$ may fall yet further.

\begin{figure}
\plotfiddle{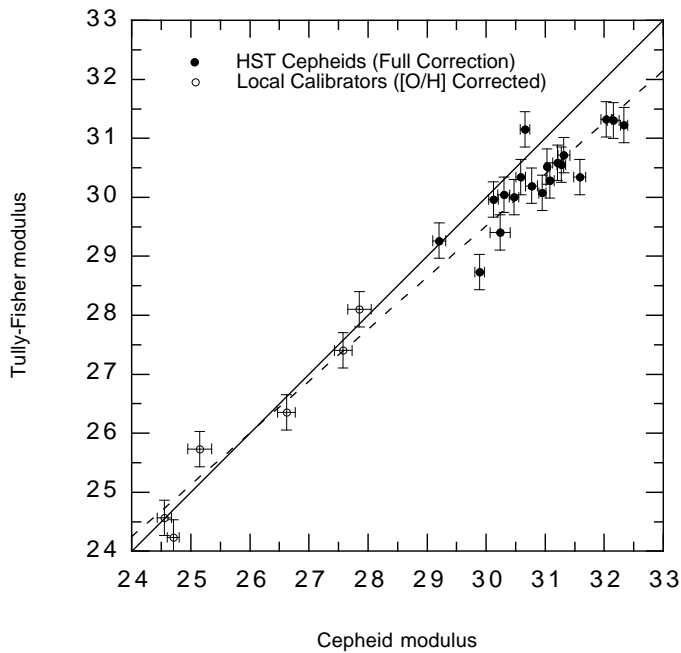}{3in}{0}{75}{75}{-220}{-305}
\caption{Tully-Fisher versus metallicity/incompleteness
corrected HST Key Project Cepheid distances (Allen \& Shanks, 2004).
The TF relation  underestimates Virgo galaxy distances
by 34$\pm$6\%. The least squares fit (dashed line) shows 
3.5$\sigma$ evidence for a TF scale error.} 
\end{figure}

Some 25 galaxies have had Cepheids detected by HST. Seventeen of these
were observed by the HST Distance Scale Key Project (Freedman et al.,
1994, Ferrarese et al., 2000). Seven were observed in galaxies with
SNIa by Sandage and collaborators (eg Sandage et al., 1996) and M96 in
the Leo I Group was observed by Tanvir et al. (1995). Allen \& Shanks
(2004) have used these data to update the comparison of I-band
Tully-Fisher (TF) distances of Pierce \& Tully (1992) with the
published HST Cepheid distances. These authors find that TF distance
moduli at the Virgo distance are underestimates by $\approx$22$\pm$5\%. If the Key
Project metallicity correction (see also Hoyle, Shanks \& Tanvir,
2003) and the P-L incompleteness correction of Allen \& Shanks is
applied to the Cepheids then the TF moduli at the Virgo distance are now
underestimates by $34\pm6$\% (see Fig. 3). This reduces Tully-Fisher
estimates of H$_0$ from $\approx$85 to $\approx$65kms$^{-1}$Mpc$^{-1}$
(Giovanelli et al., 1997, Shanks 1997, Shanks, 1999, Sakai et al.,
1999).  Of course, $H_0$ might be further reduced if the TF scale
error persists to Coma. The correlation of Cepheid residuals with
line-width suggests TF distances may be Malmquist biased - possibly
implying a bigger TF scale error at larger distances. This clear
problem for TF distances, which previously has been the `gold
standard' of secondary distance indicators, warns that errors in the
extragalactic distance scale may still be seriously underestimated!

\begin{figure}
\plotfiddle{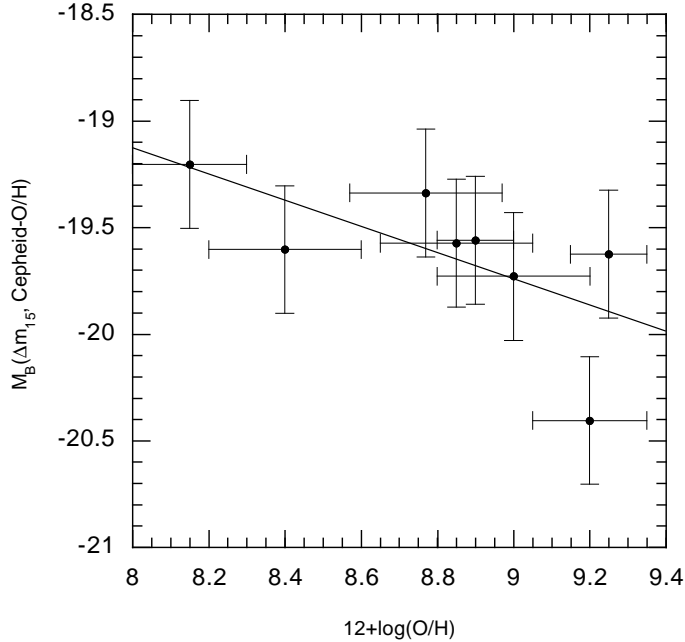}{3in}{0}{75}{75}{-220}{-305}
\caption{The SNIa absolute magnitude-metallicity relation using the SNIa
peak magnitudes of Gibson et al. (2000), now corrected for $\Delta m_{15}$  and
Cepheid metallicity/incompleteness (Allen \& Shanks 2004). The least squares
fit (solid line) shows 2$\sigma$ evidence for a correlation.} 
\end{figure}

Eight HST Cepheid galaxies also have SNIa distances. Correcting the
Cepheid scale for metallicity and incompleteness bias after Allen \&
Shanks and then using these distances to derive peak luminosities using
the SNIa data from Gibson et al. (2000) implies a possible correlation
between Type Ia peak luminosity and metallicity (see Fig. 4). Such a
scatter in SNIa luminosities could easily be disguised by magnitude
selection (Malmquist) effects at moderate redshifts. At higher redshift
the correlation is in the right direction to explain away the need for a
cosmological constant in the Supernova Hubble Diagram results, since
galaxies at high redshift might be expected to have lower metallicity.
Thus the conclusion is that if Cepheids have strong metallicity
dependence then so have SNIa and therefore SNIa estimates of q$_0$ and
H$_0$ may require  significant correction.

\section{Conclusions}

Our main conclusions are as follows:-

\begin{itemize}

\item $\Lambda$CDM gains strong support from the WMAP and Boomerang CMB
peaks and also the SNIa Hubble diagram - but leaves a standard model
which is  fine-tuned to the almost impossible level of one part in
$10^{102}$ and based on two pieces of undiscovered physics, dark energy
and cold dark matter.

\item The size of the vacuum energy density implied by the SNIa Hubble diagram
is so small that it is unstable to quantum corrections.

\item Superstring models of quantum gravity which invoke compactified
higher spatial dimensions are broadly incompatible with the positive
cosmological constant of the $\Lambda$CDM model and prefer models with negative 
or no cosmological constant.

\item $\Lambda$CDM also has astrophysical problems predicting galaxy
mass profiles that are too cuspy at small scales and a galaxy luminosity function that
is too steep. The model also has a problem with new results from QSO
lensing  that prefer a value of $\Omega_m\approx1$.

\item The main escape routes to other models include the expectation
that the SNIa Hubble diagram may require evolutionary corrections.
Further, the precision of the CMB power spectrum may still be
compromised by foreground contamination from the epoch of reionisation
at $z\approx15$ and the SZ signal from galaxy clusters at $z\la1$.

\item We have argued that if  $H_0\la50\,$kms$^{-1}$Mpc$^{-1}$ then it might allow 
a simpler, inflationary  model with $\Omega_{baryon}=1$ and with no
need to invoke dark energy or cold dark matter.

\item The strong scale error between HST Cepheid and TF distances and
the potential metallicity dependencies for the maximum luminosity of
SNIa and the Cepheid P-L relation suggests that there may still be
systematic errors in the distance scale which may allow a significantly
lower value of $H_0$; our very simple model with $\Omega_{baryon}=1$
may therefore still not be ruled out.

\end{itemize}

Finally, we note that the fundamental weaknesses of the standard model
make the conclusion that the Universe is CDM and dark energy dominated
also vulnerable to the new higher-dimensional `brane-world'
cosmologies motivated by string theories (Randall \& Sundrum 1999,
Dvali, Gabadadze \& Porrati, 2000, Freese \& Lewis, 2002). These
cosmologies offer a rich, new variety of terms to add to the standard
Friedmann solution of the field equations. The resulting increased
flexibility in observational cosmology will at least increase the
chance of finding alternative cosmologies to $\Lambda$CDM. For
example, there exist Cardassian models that fit the current CMB and
SNIa data, assuming a baryon-dominated model. This model is still
highly finely-tuned but no more than $\Lambda$CDM. Thus whether the
increased flexibility in observational cosmology arises from this
route or from the presence of systematic errors in the current
cosmological data as argued here, it seems likely that a more
satisfactory model than $\Lambda$CDM will at some stage appear and
therefore that the rumours of the `end of cosmology' may well be
premature!

\end{document}